\documentclass[a4paper,fleqn]{cas-dc}

\usepackage[numbers]{natbib}

\def\tsc#1{\csdef{#1}{\textsc{\lowercase{#1}}\xspace}}
\tsc{WGM}
\tsc{QE}

\begin{document}
\let\WriteBookmarks\relax
\def\floatpagepagefraction{1}
\def\textpagefraction{.001}
\let\printorcid\relax 

\shorttitle{Energy consumption optimization and self-powered environmental monitoring design for low-carbon smart buildings}    

\shortauthors{Zhengbao Yang et al.}

\title[mode = title]{Energy consumption optimization and self-powered environmental monitoring design for low-carbon smart buildings}

\author[1]{Yuhan Dai}

\author[1]{Mingtong Chen}

\author[1]{Zhengbao Yang}
\cormark[1] 
\ead{zbyang@hk.ust} 
\ead[URL]{https://yanglab.hkust.edu.hk/}

\address[1]{The Hong Kong University of Science and Technology
Hong Kong, SAR 999077, China}

\cortext[1]{Corresponding author} 

\begin{abstract}
Despite the growing emphasis on intelligent buildings as a cornerstone of sustainable urban development, significant energy inefficiencies persist due to suboptimal design, material choices, and user behavior. The applicability of integrated Building Information Modeling (BIM) and solar-powered environmental monitoring systems for energy optimization in low-carbon smart buildings remains underexplored. Can BIM-driven design improvements, combined with photovoltaic systems, achieve substantial energy savings while enabling self-powered environmental monitoring? This study conducts a case analysis on a retrofitted primary school building in Guangdong, China, utilizing BIM-based energy simulations, material optimization, and solar technology integration. The outcomes reveal that the proposed approach reduced annual energy consumption by 40.68\%, with lighting energy use decreasing by 36.59\%. A rooftop photovoltaic system demonstrated a payback period of 7.46 years while powering environmental sensors autonomously. Hardware system integrates sensors and an ARDUINO-based controller to detect environmental factors like rainfall, temperature, and air quality. It is powered by a 6W solar panel and a 2200 mAh/7.4 V lithium battery to ensure stable operation. This study underscores the potential of BIM and solar energy integration to transform traditional buildings into energy-efficient, self-sustaining smart structures. Further research can expand the scalability of these methods across diverse climates and building typologies.

\end{abstract}




\maketitle

\section{Introduction}

As intelligent building systems become intricately intertwined with human life, people are gradually paying attention to the benefits of intelligent buildings. Intelligent Building refers to the intelligent management of building equipment and environment through the integration of modern communication, computer, control and other technologies, in order to provide a safe, efficient, comfortable and convenient living and working environment\cite{1}. Its main systems include Office Automation System (OAS), Communication Automation System (CAS) and Building Automation System (BAS).
The development of intelligent buildings can not only improve energy utilization efficiency\cite{2} and enhance the quality of living and working environments\cite{3}; It can also promote energy conservation and emission reduction, and support the sustainable development of cities in the long term\cite{4}.
The World Economic Forum points out that buildings account for 30-50\% of global building energy consumption. In the United States, this proportion is as high as 39\%, higher than that in industry and transportation. 75\% of the energy used by buildings is for HVAC and lighting.
However, in actual energy usage, a large amount of energy is wasted due to reasons such as the thermal bridge effect\cite{5} and insufficient insulation performance of building envelopes\cite{6}, improper user behavior and management\cite{7}.
In terms of energy conversion and utilization, depending on the type of energy, the mainstream energy technologies at home and abroad currently include offshore wind power technology\cite{8}, hydropower technology\cite{9}, geothermal system technology\cite{10} and solar cell technology\cite{11}. Among them, solar cells have attracted much attention due to their high efficiency and low cost, and have developed rapidly in recent years. Compared with other renewable energy sources such as wind energy, hydropower and geothermal energy, solar energy is abundant in resources, has a wide range of applications, and the corresponding technical equipment is flexible to instal. Moreover, with the development of photovoltaic materials and manufacturing processes, its cost has dropped rapidly in recent years\cite{12,13}.
In current architectural design, the methods for analyzing energy consumption and optimizing design mainly include Building Energy Consumption Simulation (BEM)\cite{14}, artificial intelligence and machine learning\cite{15}, and the integration of Building Information Modeling (BIM)\cite{16}, etc.  BIM technology has several significant advantages over traditional methods in building energy analysis and optimization. It allows for the integration of energy performance analysis in the early stage of architectural design, promoting collaborative work among architects, engineers and other stakeholders\cite{17}. And it is integrated with real-time energy consumption simulation tools, enabling designers to quickly evaluate the energy performance of different design schemes at an early stage\cite{18}.
In today's field of intelligent buildings, environmental monitoring sensors are often installed inside buildings to conduct real-time monitoring of temperature, humidity, smoke, fire and other conditions inside the buildings. The data collected by the sensors is transmitted to the cloud via wireless communication or decisions are made directly at the edge layer\cite{19}. This greatly ensures the safety of the building and its users, and significantly enhances the convenience of building maintenance. The integration of various sensors makes buildings increasingly "intelligent", turning them into buildings with "cognition"\cite{20}.
This project mainly adopts BIM technology, supplemented by lighting simulation analysis software, to conduct energy analysis and optimization of the selected teaching building in a specific area, making the renovated teaching building more intelligent. The goal is to achieve an average annual energy-saving rate of at least 30\% through the use of green energy and design improvements. The main specific optimization parameters include at least 50\% energy savings on average annually in lighting and at least 30\% energy savings on average annually in the envelope structure. Solar cell-powered environmental monitoring system integrates sensors and an ARDUINO-based controller to detect environmental factors like rainfall, temperature, and air quality. It is powered by a 6 W solar panel and a 2200 mAh/7.4 V lithium battery to ensure stable operation. And explore the feasibility of a multi-environment monitoring system powered by solar cells.

\section{Energy analysis and optimization}

Analysis requires the selection of a specific building in reality to construct its digital twin model, and the application of software to better understand its relevant information and take optimization measures. Considering the best applicable scenarios for solar cells, we selected Guangdong Province, China, which enjoys abundant sunshine throughout the year. So, the teaching buildings in the school are taken as the main building types for analysis.
\begin{figure}[h]
	\centering
		\includegraphics[scale=1]{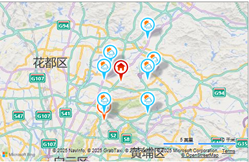}
	  \caption{The selected location is in Guangdong Province}\label{FIG:1}
\end{figure}

Ultimately, an old teaching building in a primary school located in a certain area of Guangdong Province was selected as the experimental target. This building is a three-story detached structure with a main entrance facing northeast, two evacuation passages and a atrium. There are a total of 16 classrooms and multiple multi-functional activity rooms and equipment rooms. The building is directly restored in a 1:1 parametric manner in the BIM modeling system. 

\begin{figure}[h]
	\centering
		\includegraphics[scale=1]{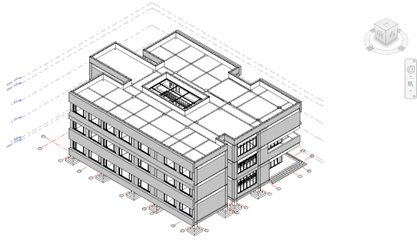}
	  \caption{A 1:1 restored model in BIM}\label{FIG:2}
\end{figure}

After improvement, the overall system is a typical water system +VAV terminal air conditioning system, which is applied in the teaching building scene. Its core components are: The Condensing Water system (CDW01) supports the operation of the chiller. The chilled water system (CHW01) provides chilled water and is connected to AHU01 for air cooling. The Low Pressure Hot Water system (LP HW 01) provides low-pressure hot water to heat the variable air volume box (VAV01 / VAV02). AHU01 is the core equipment for air handling and supplies air to VAV01 / VAV02.

According to software simulation, the annual energy consumption of the current building is as follows: The total annual energy consumption is 1,664.82 GJ, among which lighting accounts for 10.23\% of the terminal usage energy consumption(170.33 GJ), and the cooling and heating loads account for 68.94\% and 7.52\% respectively. Other equipment accounts for 13.3\%.The trend chart of the peak demand for cooling and heating loads throughout the year varying with time is as follows. It can be seen that it varies significantly with the seasons in Guangdong Province. The demand for heating in winter is not as great as that for cooling in summer.

Insight is a cloud service integrated with Revit, using Revit's energy analysis model as the starting point for operational (OC) and embodied carbon (EC) analysis. First, modules are selected for the space and surface of the building, including architectural elements (walls, floor slabs, roofs, etc.). The Spaces and surfaces of these envelope structures are the key paths for energy exchange in buildings. As shown in the following figure, the dark blue surface represents the floor and roof components, the green surface represents the wall components, the light blue surface represents the window glass components, and the gray surface represents other components such as railings and stairs.

\begin{figure}[h]
	\centering
		\includegraphics[scale=1]{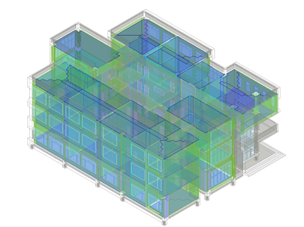}
	  \caption{Building space and surface}\label{FIG:3}
\end{figure}

After the energy model is created, it can be analyzed and optimized in Insight and visual results can be generated. By setting the initial values of all aspects, it can be obtained that the original energy consumption of the building is 177 kWh/m² / yr. Then, by changing the values of different influencing factors within a certain range, the images of EUI (Energy Use Intensity, The unit is kWh/m² / yr) varying with different influencing factors can be obtained.

Firstly, explore the influence of window-to-wall ratio (WWR) in different orientations of buildings on energy consumption to carry out possible optimized designs. It can be seen that in all four orientations, reducing the window-to-wall ratio can always lower energy consumption. However, considering factors such as design aesthetics and lighting and ventilation, the selection of window-to-wall ratios in different orientations should be based on reality.
Guangdong Province is a region with hot summers and warm winters. In summer, the solar radiation is intense, and the air conditioning load mainly comes from the heat gained from solar radiation. Therefore, the window-to-wall ratio needs to be strictly controlled to reduce energy consumption. According to the "Energy Efficiency Design Standard for Public Buildings" (GB 50189-2005), in hot summer and warm winter areas, the shading coefficient of exterior Windows should be given priority consideration, and limit values should be set for the window-to-wall ratio in different orientations at the same time. South-facing Windows usually allow for higher window walls, such as 0.4-0.7, to take advantage of natural lighting, but shading measures must be taken in conjunction. Due to the high intensity of solar radiation in the east-west direction, the window-to-wall ratio is strictly limited, such as <0.35, to avoid excessive heat gain. The north-facing lighting demand is relatively low, and the window-to-wall ratio can be appropriately reduced, such as <0.45.
After optimization, in order to enhance lighting, the WWR for the southbound direction was increased from 24\% to 40\%-65\%. Subsequently, shading design can be combined to reduce energy consumption. The WWR for other orientations all meet the requirements and no major changes are needed.

The current building has not adopted any shading measures, so an optimized design is needed. It can be concluded from the image that changing the proportion of the sun visor facing south has the most obvious effect. The northbound and eastbound images overlap.
In summer, the solar altitude Angle in Guangdong Province is high (about 87°). Horizontal sunshades need to cover direct sunlight, but excessive shading of winter sunlight should be avoided (the altitude Angle on the winter solstice is 43°). It is recommended that the proportion of south-facing shading height be between 1/3 and 1/2. If the proportion is too low (such as 1/6), the shading effect will be insufficient. If the proportion is too high (such as 2/3), it may affect natural lighting and require additional supplementary lighting, increasing energy consumption. Actual cases (such as a certain middle school in Shenzhen) show that a 1/3-1/2 proportion combined with Low-E glass (SC<0.4) can reduce energy consumption by 20\% - 30\%. For east-west windows, it is recommended to have a shading height ratio of 1/4 to 1/3. For north-facing windows, the shading height ratio can be appropriately reduced, if it is no more than 1/4, to mainly allow natural lighting.

Below, we will respectively explore the impact on building energy consumption from the aspects of window glass material, wall construction material and roof construction material.For window glass, single-layer ordinary transparent glass (SGL Clr) is currently adopted. This type of glass has poor energy-saving performance, no thermal insulation, and high energy consumption throughout the year. After comparison, taking into account regional characteristics, cost and energy conservation, double-layer low-emissivity coated glass (Dbl LoE) is adopted. It has good heat insulation performance and the best cost performance. Although double-layer ordinary transparent glass (Dbl Clr) is similar to LoE glass in terms of energy efficiency and is cheaper, it is not recommended for use because it has no sun protection and performs poorly in areas with strong sunlight.

For the wall material, the current buildings use walls without insulation layers. After comparison, it is recommended to use a 12.25-inch thick structural insulating board with a thermal resistance of approximately R10 to R12. This material has good heat insulation, is easy to construct, and has a lower cost compared to R14-inch ICF.For the roof material, the current buildings use concrete with a thermal resistance of 2, which has poor thermal insulation and heat preservation performance. After comparison, it is recommended to adopt a structural insulating board with a thickness of 10.25 inches.Regarding permeability, this building is an old one with high permeability. During the improvement construction stage, using materials with high air tightness and strengthening the joint treatment can significantly enhance the building's air tightness and reduce the permeability to 0.4-0.8ACH.In Insight, the changes in energy consumption after altering the lighting efficiency of artificial light sources and replacing the air conditioning system can be roughly estimated initially. The original building adopted high-power incandescent lamps and VAV air conditioning systems. After improvement, the adoption of energy-saving LED lights and high-efficiency heat pump technology can significantly reduce energy consumption. Heat pump technology has a lower operating cost than VAV and is a more advanced energy technology.

After the above improvements, the EUI of the teaching building has decreased from 177 kWh/m² / yr to 105 kWh/m² / yr. Meet the ASHRAE 90.1 energy standard (269 kWh/m² / yr). The next step is expected to achieve the carbon neutrality standard of Building 2030 (72 kWh/m² / yr).Next, we calculate the reduced cost. According to the charging standards of electricity and gas fees in Guangdong Province, the electricity fee is CNY0.66 per Kwh and the gas fee is CNY3.41 per cubic meters. Then the average annual cost decreased from 102.8 CNY/m²/yr to 71.7 CNY/m²/yr.

The lighting analysis of the built environment is divided into natural light and artificial light. Use the lighting analysis of BIM to simulate the natural lighting conditions (i.e., the optimal lighting time) of a certain morning in summer in this area. The results show that the atrium has good lighting and there is no need to supplement too many artificial light sources during the day. In the classroom, there is insufficient light and low brightness at a distance from the Windows. Artificial light sources need to be supplemented.

\begin{figure}[h]
	\centering
		\includegraphics[scale=0.85]{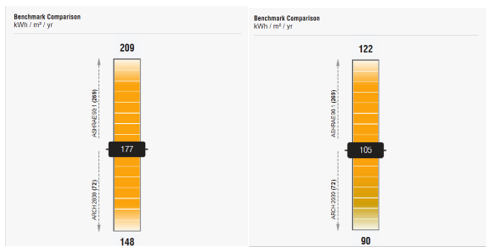}
	  \caption{Left image: EUI before optimization, right image: EUI after optimization}\label{FIG:4}
\end{figure}

\begin{figure}[h]
	\centering
		\includegraphics[scale=0.85]{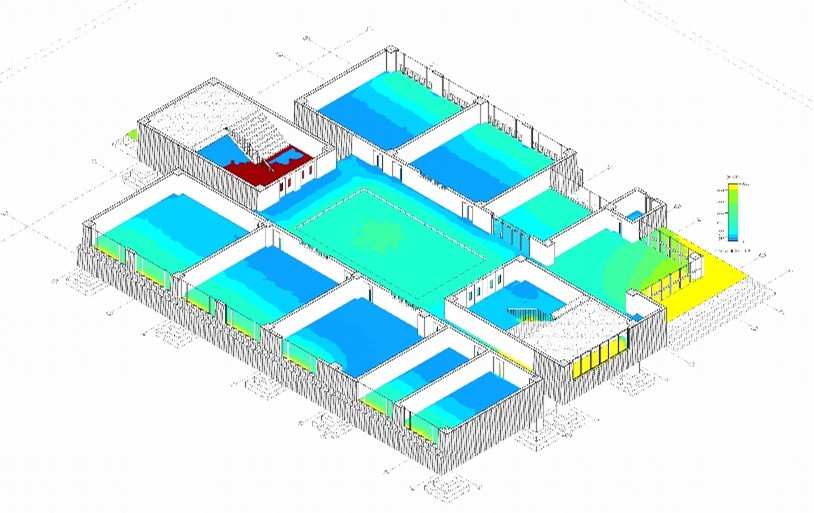}
	  \caption{1 F Natural light distribution}\label{FIG:5}
\end{figure}

\begin{figure}[h]
	\centering
		\includegraphics[scale=0.85]{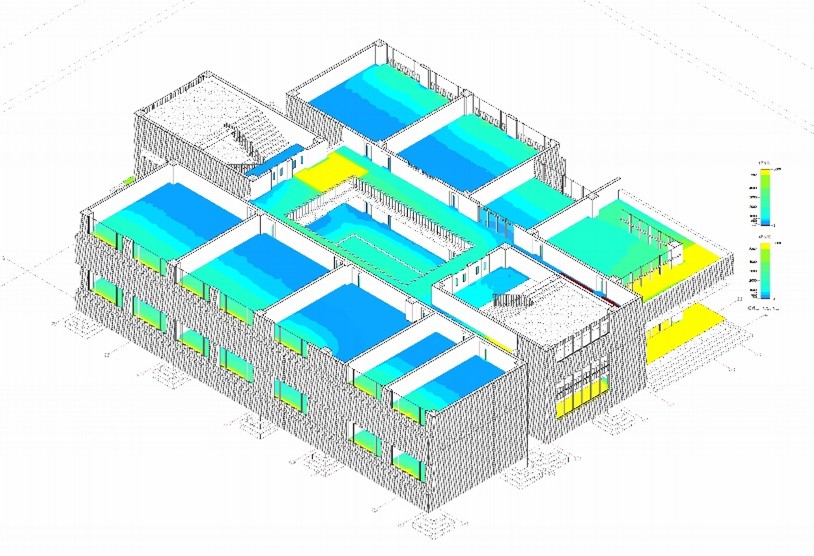}
	  \caption{2 F Natural light distribution}\label{FIG:6}
\end{figure}

\begin{figure}[h]
	\centering
		\includegraphics[scale=0.85]{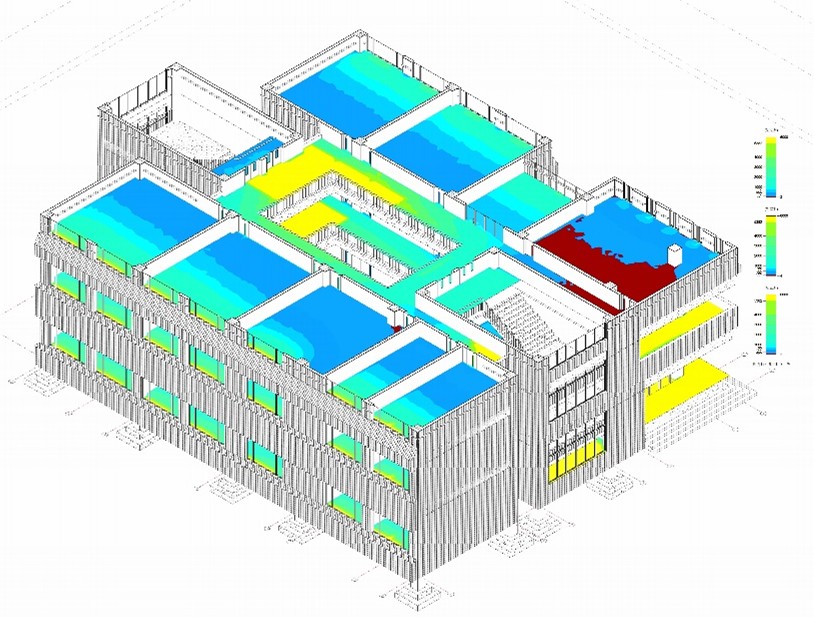}
	  \caption{3 F Natural light distribution}\label{FIG:7}
\end{figure}

The lighting analysis software can assist designers in selecting lamps for the interior environment of specific buildings and making layout presets, with the aim of achieving a good lighting environment. The lighting in the classroom should be evenly distributed and have an illuminance of > 500 lux. On the lighting analysis, the selection and arrangement of artificial lamps are carried out to achieve the expected effect and the required amount of light sources and total energy consumption are calculated. After calculation, when using a linear lamp arrangement, approximately 500 lamps are needed in total. The total annual energy consumption is approximately 108.01 GJ.

Intelligent buildings require multiple sensors to collect data in real time. Introduce new sensors to the teaching buildings that need improvement, such as temperature and humidity sensors and air quality sensors to monitor the environment in real time, and access control card sensors to ensure the safety of students, etc. The following table shows the types and specification parameters of sensors that can be installed in teaching buildings and are collected from the market.Based on the number and power of the sensors, the total annual energy consumption of all sensors in the building can be calculated to be 280kw. According to the electricity fee standard of Guangdong Province, the annual electricity fee is approximately CNY185.

Introducing photovoltaic systems to teaching buildings has multiple advantages. First of all, Guangdong Province enjoys a superior geographical location and abundant sunshine throughout the year. Secondly, the photovoltaic system is clean and pollution-free, which is conducive to the creation of green schools. Finally, it has long-term and objective economic benefits and is supported and subsidized by the government.
Calculating the actual investment payback period can better predict the economic effects of installing photovoltaic systems. The available area of the roof of the teaching building is approximately 700 square meters. Select solar panels available on the market that are 1.6 meters long and 1 meter wide, with a power of approximately 300w. If the roof is fully covered and a maintenance passage and safety distance are reserved, approximately 350 pieces can be installed. The total capacity is approximately 105kW. Based on the sunshine duration in Guangdong Province, it can be estimated that the average annual equivalent power generation is 1,200 hours. Then, the annual power generation = 105 × 1200 = 126,000 kWh. The annual electricity consumption of the teaching building is approximately 30,283 KWh (lighting +sensor), and the annual electricity bill is about CNY 19,986.78. Residual electricity = 126,000-30,283 = 95,717 KWH, annual savings + benefits = 19,986.78 + 95,717 *0.35= CNY53,487.73. The market installation price is approximately CNY3.8/W, the investment cost = CNY105,000 W*3.8= CNY399,000, and the payback period = 399,000/53,487.75 =7.46 years.


\section{Solar cell-powered environmental monitoring system}

In the above article, the layout of photovoltaic systems and sensors is of vital importance to smart buildings. Next, we will explore the feasibility of green energy power supply.
Connect a system used for detecting the environment and responding. The system composition includes the sensor part and the controller part. Sensors are used to detect rainfall, temperature and humidity, air quality, etc.The controller is based on ARDUINO to achieve data acquisition and control. The power supply scheme adopts 165 mm×195 mm solar panels (rated voltage 12 V, rated current 500 mA, peak power 6 W). Equipped with a 2200 mAh/7.4 V conversion lithium battery, it ensures the stable operation of the system. After connection, the equipment was taken to a sunny place. Experiments showed that the overall operation was good.When raindrops fall on the sensor and the rainfall reaches the preset threshold, the alarm light will come on.

\begin{figure}[h]
	\centering
		\includegraphics[scale=1]{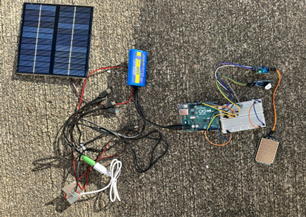}
	  \caption{Composition of experimental equipment}\label{FIG:8}
\end{figure}

\begin{figure*}[h]
	\centering
		\includegraphics[scale=0.8]{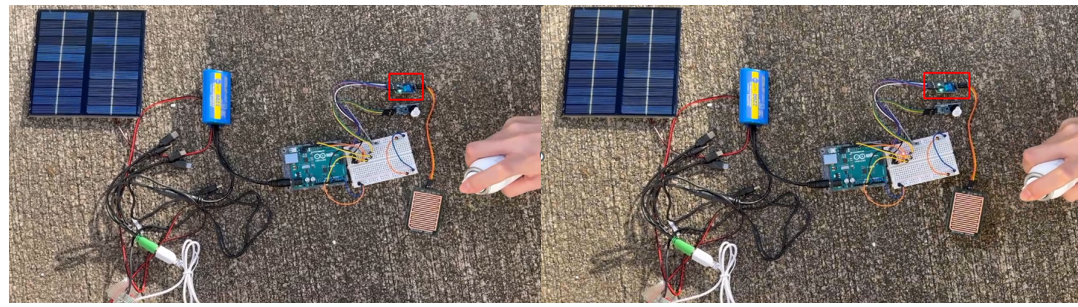}
	  \caption{The left: before watering, the right: after watering}\label{FIG:9}
\end{figure*}

\section{Discussion and Conclusion}

Through design optimization, the annual average energy consumption has decreased from 177 kWh/m² / yr to 105 kWh/m² / yr, and the energy-saving rate has decreased by 40.68\%. The annual energy consumption of lighting decreased from the original 170.33 GJ to 108.01 GJ, a reduction of 36.59\%. Due to the introduction of the photovoltaic system, after 7.46 years, ideally, an additional annual income of CNY 33,500.95 can be generated. The influencing factors of the envelope structure include WWR, shading, window glass, etc. After the transformation, the average annual energy consumption is reduced by 29 kWh/m²/yr, accounting for 40.28\% of the total energy reduction.
The success of the exploration experiment of raindrop sensors powered by solar cells means that green energy can be widely applied in intelligent building environment monitoring systems, including the monitoring of temperature, humidity and smoke. It has a very good development prospect and profound significance. In the long run, this integration of clean energy with smart sensing technologies not only reduces carbon emissions and operational costs, but also lays a foundation for the realization of fully self-sustaining, low-maintenance smart campuses and urban infrastructure.










\bibliographystyle{cas-model2-names}

\bibliography{cas-refs}



\end{document}